\documentclass{JHEP3}
\newcommand{\beq}{\begin{eqnarray}}
\newcommand{\eeq}{\end{eqnarray}}

\def\be{\begin{equation}}
\def\ee{\end{equation}}
\def\bea{\begin{eqnarray}}
\def\eea{\end{eqnarray}}

\title{ Correlation functions in a cascading N=1 gauge theory}
\author{Michael Krasnitz\\ Joseph Henry Laboratories,
Princeton University,
Princeton, New Jersey 08544\\ E-mail: \email{krasnitz@princeton.edu}}
\abstract{We study fluctuations around the warped conifold supergravity solution of Klebanov and Tseytlin \cite{kt}, known to be dual to a cascading ${\cal N}=1$ gauge theory. Although this supergravity background is not asymptotically AdS, corresponding to a non-conformal field theory, it is possible to apply the usual methods of AdS/CFT duality to extract the high energy behavior of field theory correlators by solving linearized equations of motion for fluctuations around the background. We consider the Goldstone vector dual to the anomalous R-symmetry current and compute its mass, which exactly matches the general prediction of \cite{bdfp}. We find the high energy 2-point functions for the R-current and two other vectors. As expected, the R-current 2-point function has a longitudinal part because R-symmetry is broken. We also calculate the high energy 2-point function of the energy-momentum tensor from fluctuations of modes in the graviton sector. This 2-point function has a trace part corresponding to broken conformal symmetry.}
\preprint{PUPT-2049} 
\begin{document}
\def\fixit#1{}
\def\comment#1{}
\def\equno#1{(\ref{#1})}
\def\equnos#1{(#1)}
\def\sectno#1{section~\ref{#1}}
\def\figno#1{Fig.~(\ref{#1})}
\def\D#1#2{{\partial #1 \over \partial #2}}
\def\df#1#2{{\displaystyle{#1 \over #2}}}
\def\tf#1#2{{\textstyle{#1 \over #2}}}
\def\Leff{L_{\rm eff}}
\def\l{\log(r/r_0)}
\def\ls{\log^2(r/r_0)}
\def\dB{\delta B}
\def\g{\gamma}
\def\p{\partial}
\def\d{\delta}
\def\P{\prime}
\def\m{\mu}
\def\n{\nu}
\def\r{\rho}
\def\s{\sigma}
\def\gr{\gamma_{rr}}
\def\f{\varphi}
\def\dP{\delta\Phi}
\def\t{\tilde}
\def\c{\cdot}
\def\sqr#1#2{{\vcenter{\vbox{\hrule height.#2pt
         \hbox{\vrule width.#2pt height#1pt \kern#1pt
            \vrule width.#2pt}
         \hrule height.#2pt}}}}
\def\square{\mathop{\mathchoice\sqr56\sqr56\sqr{3.75}4\sqr34\,}\nolimits}

\section{Introduction}

A very important aspect of the gauge theory/supergravity correspondence \cite{malda1, gkp, witt1} is that it provides an explicit procedure for calculating the full quantum gauge theory correlation functions by solving the equations of motion of classical supergravity. The classical supergravity (SUGRA) action evaluated on the classical solutions of these equations is equal to the gauge theory generating functional for connected correlation functions, with the asymptotic boundary values of SUGRA fields proportional to the sources coupling to the dual field theory operators \cite{gkp, witt1}. By taking functional derivatives of the SUGRA action with respect to these boundary values, we obtain the corresponding field theory correlators. Of course, the full nonlinear SUGRA equations of motion are generally too hard to solve, but by deriving and solving linearized equations for fluctuations around a SUGRA background we can extract at least the 2-point functions of the corresponding gauge theory operators in the field theory dual to that background.

While the original gauge/gravity correspondence was formulated for ${\cal N}=4$ Super Yang Mills (SYM) and its dual $AdS_5 \times S^5$ background, in recent years some progress has been achieved in finding new SUGRA backgrounds, dual to gauge theory models with fewer supersymmetries and nontrivial RG flow \cite{gppz,fgpw,kn,kt,ks,malda2}. Such models are hoped to be a step in the direction of finding the string duals of realistic gauge theories; indeed, there is reason to believe that the warped deformed conifold background found in \cite{ks}, as well as the similar backgrounds of \cite{malda2,vafa}, capture the correct topology of the string dual to pure ${\cal N}=1$ SYM. The conifold is warped, corresponding to a running of the coupling; deformed, corresponding to chiral symmetry breaking; and the metric in the infrared (IR) is nonsingular, allowing one to derive the area law for the Wilson loop. Because of these promising features, it is of interest to further investigate such models.

In a previous paper \cite{kras}, we took a first step towards extending the program of computing field theory correlators from SUGRA fluctuations to one of these models, the cascading ${\cal N}=1$ SUSY gauge theory and its SUGRA dual, the Klebanov-Strassler (KS) solution \cite{ks}. Using as an example the minimally coupled massless scalar, we developed a procedure for extracting the high energy behavior of 2-point functions by solving fluctuation equations around the ultraviolet (UV) limit of the KS solution, the Klebanov-Tseytlin (KT) solution \cite{kt}. The KT metric develops a singularity in the IR, resolved in the full KS background, but the UV behavior of correlators is insensitive to the details of singularity resolution. Using the simple UV cutoff regularization method of \cite{gkp}, along with an effective IR cutoff of the KT metric, we were able to obtain a sensible 2-point function for the dimension 4 operator dual to the minimal massless scalar.

In this paper, we use the method developed in \cite{kras} to calculate new correlation functions from the KT background. In particular, we consider the SUGRA modes dual to the gauge-theory R-current $J^R_{\m}$ and the gauge theory energy-momentum tensor $T_{\m\n}$. By solving the appropriate fluctuation equations, we are able to extract the leading high-energy behavior of the 2-point functions $\langle J^R_{\m}(k)J^R_{\n}(-k) \rangle$ and $\langle T_{\m\n}(k)T_{\r\s}(-k) \rangle$. Because R-symmetry and conformal symmetry are broken, these correlators are expected to have longitudinal and trace parts, respectively, in addition to the transverse parts present in CFT. Indeed, these parts are found. Because the operators $J^R_{\m}$ and $T_{\m\n}$ belong to the same supermultiplet (the supercurrent), their 2-point functions should be related to each other by supersymmetry Ward identities \cite{f3}. While we do not check these identities in detail, the form of the correlators we find suggests that they are in fact satisfied.

An important feature of the KT solution that distinguishes it from, e.g., the RG flow backgrounds investigated in \cite{bdfp}--\cite{f3} is that the KT metric is not asymptotically AdS, meaning that the dual gauge theory is non-conformal at arbitrarily high energies. As a result, the extreme UV behavior of correlation functions is nontrivial, and that is indeed what we will be interested in. In contrast, the geometries studied in \cite{bdfp}--\cite{f3} are asymptotically AdS in the UV; the breaking of conformal and R-symmetries are IR phenomena, achieved by either adding relevant operators to the AdS Lagrangian, or turning on VEVs of scalar fields. This fact gives our work a different character from that of \cite{bdfp}--\cite{f3}. The authors of those papers were concerned with the IR behavior of correlation functions, since in the models they considered the UV behavior is conformal. They found that to obtain a sensible IR behavior with the correct pole structure in momentum space, they needed to go beyond the naive AdS cutoff regularization \cite{f1,aft} and develop a holographic renormalization scheme \cite{f2,f3} (for a review, see \cite{kostas}). This scheme involves adding covariant counterterms to the regularized SUGRA action to cancel divergent contact terms. In the UV, these counterterms do not change the qualitative behavior of ``naive'' 2-point functions obtained by simply throwing out the contact terms, though they may renormalize the numerical coefficients. Since in our paper we are concerned with the UV behavior of correlators, and since holograhic renormalization does not qualitatively change that behavior, we will not be careful about including these covariant counterterms. Moreover, in one case of interest (the minimal massless scalar) we will argue that the numerical prefactor is not renormalized. In general, though, one must include covariant counterterms to obtain the exact correlators\footnote{I am indebted to Kostas Skenderis for pointing this out to me.}. Another feature of the present work is that unlike in \cite{f1,aft,f2,f3}, the fluctuation equations we derive are not exactly solvable, but we develop an iteration and matching procedure that allows us to extract the leading high-energy behavior of the correlators.

The plan of the paper is as follows. In Section 2 we review the KT solution. In section 3 we review the method developed in \cite{kras} for extracting 2-point functions from fluctuation equations around the KT background. In section 4 we derive fluctuation equations for a family of modes including the Goldstone vector dual to the R-current operator and use our method to obtain the R-current, as well as other, 2-point functions. In Section 5 we repeat this procedure for the graviton sector and compute the EM tensor 2-point function. In section 6 we discuss our results and some open questions.

Before we proceed, let us establish notation. Whenever we are in Minkowski space, the signature is $(-,+,\ldots,+)$; we will not be too careful about distinguishing Minkowski from Euclidean space. We will use Greek indices $\m,\n\ldots$ for flat 4-dimensional (Minkowski or Euclidean) space, lowercase Latin indices $i,j\ldots$ for the 5-dimensional space of noncompact dimensions ($x^{\m},r$) and uppercase Latin indices $M,N\ldots$ for the full 10-dimensional space. The operator $\square$ will denote the 4-dimensional flat space d'Alembertian, $\square_{10}$ the full 10-dimensional Laplacian. $h$ will always be used to denote the metric warp factor $h(r)$; we will often suppress the $r$-dependence.
  
\section{The Klebanov-Tseytlin solution}

In this section we review the KT solution. Our starting point is the type IIB supergravity action in the Einstein frame
$$S = -{1 \over 2\kappa^2}\int d^{10}x\lgroup \sqrt {-g}[R-{1 \over 2}(\partial \Phi)^2-{1 \over 2}e^{2\Phi}(\partial C)^2-{1 \over 12}g_se^{-\Phi}H_3^2-{1 \over 12}g_se^{\Phi}{\tilde F_3}^2-{1 \over 4 \cdot 5!}g_s^2{\tilde F_5}^2]-g_s^2C_4 \wedge F_3 \wedge H_3 \rgroup,$$
where 
\begin{eqnarray*}
\label{formdefs}
F_3 = dC_2, & H_3 = dB_2, & F_5=dC_4, \\ 
{\tilde F_3}=F_3-CH_3, & {\tilde F_5} = F_5+B_2 \wedge F_3.
\end{eqnarray*}
Here $C$, $C_2$ and $C_4$ are the RR 0, 2 and 4-forms respectively, $B_2$ is the NS-NS 2-form and $\Phi$ is the dilaton. The gravitational coupling constant $\kappa$ is related to the string tension $\alpha^{\P}$ and string coupling $g_s$ by $\kappa=8\pi^{7/2}g_s(\alpha^{\P})^2$. The resulting equations of motion are:
\begin{eqnarray}
\label{einstein}
R_{MN} = {1 \over 2}\partial_M\Phi\partial_N\Phi+{1 \over 2}e^{2\Phi}\partial_MC\partial_NC+{1 \over 96}g_s^2{\tilde F_{MPQRS}}{\tilde F_N^{\; PQRS}}+ \nonumber \\
+{g_s \over 4}(e^{-\Phi}H_{MPQ}H_N^{\; PQ}+e^{\Phi}{\tilde F_{MPQ}}{\tilde F_N^{\; PQ}})- \nonumber \\
-{1 \over 48}g_{MN}(e^{-\Phi}H_{PQR}H^{PQR}+e^{\Phi}{\tilde F_{PQR}}{\tilde F^{PQR}}),
\end{eqnarray}
\begin{eqnarray}
\label{forms}
d\star (e^{\Phi}{\tilde F_3}) = g_sF_5 \wedge H_3, \nonumber \\
d\star (e^{-\Phi}H_3-Ce^{\Phi}{\tilde F_3}) = -g_sF_5 \wedge F_3,
\end{eqnarray}
\begin{eqnarray}
\label{scalars}
d\star d\Phi = e^{2\Phi}dC \wedge \star dC-{g_s \over 2}e^{-\Phi}H_3 \wedge \star H_3 +{g_s \over 2}e^{\Phi}{\tilde F_3} \wedge \star {\tilde F_3}, \nonumber \\
d(e^{2\Phi}\star dC) = -g_se^{\Phi}H_3 \wedge \star {\tilde F_3}.
\end{eqnarray} 
These equations are supplemented by the self-duality condition
\begin{equation}
\label{self}
\star {\tilde F_5} = {\tilde F_5}.
\end{equation}
In what follows we will also need the structure of the manifold $T^{1,1}$. This is a compact five-dimensional Einstein manifold with topology $S^2 \times S^3$; it has a nontrivial two-cycle and a nontrivial three-cycle. The coordinates on $T^{1,1}$ are the angles $\psi$, $\theta_1$, $\theta_2$, $\phi_1$, $\phi_2$ with $\theta_1$, $\theta_2$ $\in [0,\pi]$, $\phi_1$, $\phi_2$ $\in [0,2\pi]$, $\psi$ $\in [0,4\pi]$. We define the 1-forms    
\begin{eqnarray*}
\label{gforms}
g^1 = {e^1-e^3 \over \sqrt {2}}, \;   g^2 = {e^2-e^4 \over \sqrt {2}}, \\
g^3 = {e^1+e^3 \over \sqrt {2}}, \;   g^4 = {e^2+e^4 \over \sqrt {2}}, \\
g^5 = e^5,
\end{eqnarray*}
where
\begin{eqnarray*}
\label{eforms}
e^1 = -\sin \theta_1d\phi_1,  \; e^2=d\theta_1, \\
e^3 = \cos\psi\sin\theta_2d\phi_2-\sin\psi d\theta_2, \\
e^4 = \sin\psi\sin\theta_2d\phi_2+\cos\psi d\theta_2, \\
e^5 = d\psi + \cos\theta_1d\phi_1+\cos\theta_2d\phi_2.
\end{eqnarray*}
The metric on $T^{1,1}$ is
$$ds_5^2 = {1 \over 9}(g^5)^2 + {1 \over 6}{\sum_{i=1}^4} (g^i)^2.$$
The closed nonexact forms corresponding to the nontrivial cycles are
$$\omega_2 = {1 \over 2}(g^1 \wedge g^2 + g^3\wedge g^4), \; \omega_3 = g^5 \wedge \omega_2.$$
Let us briefly review the D-brane construction that leads to the gauge theory of interest. We begin with type IIB string theory on a product of flat four-dimensional Minkowski space with a cone over $T^{1,1}$. Because $T^{1,1}$ is an Einstein manifold, the cone is Ricci-flat. We place N D3-branes at the singularity of the cone; this yields a conformal $SU(N) \times SU(N)$ gauge theory with ${\cal N}=1$ supersymmetry \cite{kw}. We then wrap an additional M D5-branes over the two-cycle of $T^{1,1}$, with the remaining D5-brane directions parallel to the D3-branes. The wrapped D5-branes become fractional D3-branes that preserve supersymmetry but break conformal symmetry. The resulting theory has gauge group $SU(N+M) \times SU(N)$ and matter in the form of two chiral superfields in the $(\overline {N+M},N)$ representation and two superfields in the $(N+M,\overline{N})$ representation \cite{kn,kt}. This gauge theory undergoes a series of Seiberg duality cascades. In the infrared it flows to a confining phase with broken chiral symmetry \cite{ks}.

To obtain the supergravity dual of this gauge theory we make the usual warped ansatz for the metric:
\beq
\label{warped}
ds^2 = h^{-1/2}(r)dx_{\mu}dx^{\mu}+h^{1/2}(r)(dr^2+r^2ds_5^2).
\eeq
Also, to get the correct number of wrapped D5-branes, we require
$$F_3 = {1 \over 2}M\alpha^{\prime}\omega_3.$$
It turns out that to find a solution of the equations of motion, we must turn on the NS-NS 2-form $B_2$. As a result, the D3-brane charge cannot be kept constant and begins to flow, corresponding to the duality cascade. The full solution is given by \cite{kt}
\begin{eqnarray}
\label{ktsol}
h(r) = {R^4+2L^4(\log(r/r_0)+1/4) \over r^4}, \;
B_2 = {2L^2 \over 3}\log(r/r_0)\omega_2, \; F_3 = {2L^2 \over 9g_s}\omega_3, \nonumber \\
F_5 = (\partial_rh^{-1})d^4x\wedge dr+{R^4 \over 27}g^1 \wedge \ldots \wedge g^5, \;
\Phi=C=0,
\end{eqnarray}
where $h(r)$ is the warp factor in the metric (\ref{warped}) and the radii $R$, $L$ are given by
\begin{equation}
\label{scales}
L^2 = {9 \over 4}g_sM\alpha^{\prime}, \; R^4 = {27 \over 4}g_sN\pi(\alpha^{\prime})^2.
\end{equation}
This is the Klebanov-Tseytlin solution. Note that in addition to the string scale $\alpha^{\prime}$, the solution involves an arbitrary scale $r_0$. This scale is related to the confinement scale; we shall see the precise relation in the next section. As we flow toward the IR, the solution becomes singular at a radius $r=r_s$ where $h(r_s)=0$, and is only reliable away from the singularity, or for $r \gg r_s$. In this region the curvature $R$ satisfies $\alpha^{\prime}R \sim {1 \over g_sM}\log(r/r_0)^{-3/2} \ll 1$, so supergravity is a good approximation to the dual gauge theory. Importantly, the KT solution is {\it not} asymptotically AdS in the UV, ot at large $r$: the warp function $h$ differs from the AdS warp function by a logarithmic factor. Because the difference is only logarithmic, though, there is hope that some of the methods developed for AdS/CFT can be applied to the KT background; we will see that this is in fact the case.

A striking feature of the solution is that the D3-brane charge is scale dependent, namely
\begin{equation}
\label{cascade}
N_{eff}(r) = {1 \over (4\pi^2\alpha^{\prime})^2}\int_{T^{1,1}}{\tilde F_5} = N+{3 \over 2\pi}g_sM^2\log(r/r_0).
\end{equation}
As shown in \cite{ks}, this logarithmic running of the effective number of colors corresponds on the gauge theory side to a cascade of Seiberg duality transformations. We will not describe this cascade here since it is not directly relevant to our purposes.

\section{Field theory correlators from the KT solution} 

In an earlier paper \cite{kras} we extended the standard AdS/CFT procedure for extracting gauge theory correlation functions from supergravity to the KT background and its field theory dual, using as an example the massless scalar and its dual dimension 4 operator. We will now review that calculation and indicate how to use the same method to derive correlation functions of other field theory operators from this supergravity background.

Let us first recall how one extracts gauge theory correlation functions from the dual supergravity background in standard AdS/CFT. We follow the method of \cite{gkp}. For every SUGRA field $\phi$ there is a corresponding gauge theory operator ${\cal O}$ such that a term $W[\phi] = \int d^4x\phi(x){\cal O}(x)$ can be added to the gauge theory action. The gauge theory/SUGRA correspondence then states
\begin{equation}
\label{corr}
\langle e^{-W[\phi(x)]} \rangle = e^{-S[\phi(x)]},
\end{equation}
where $S[\phi(x)]$ is the classical SUGRA action evaluated on the field $\phi(x,r)$ that solves the supergravity equations of motion subject to the following boundary conditions: in the UV, i.e. for $r \rightarrow \infty$, $\phi(x,r)=r^{\Delta}\phi(x)$ where $\Delta$ is related to the dimension of the operator ${\cal O}$. We also require $\phi(x,r)$ be regular at the IR, i.e. for small $r$. In other words the classical SUGRA action evaluated on the classical solution $\phi(x,r)$ subject to these boundary conditions generates the connected gauge theory correlation functions of the operator ${\cal O}$.

In particular, suppose we want to calculate the two point function $\langle{\cal O}_4(x_1){\cal O}_4(x_2)\rangle$ for an operator ${\cal O}_4$ corresponding to a minimal massless scalar $\phi$ propagating in the geometry (\ref{warped}), where $ds_5^2$ is the metric on some Einstein manifold $X^5$. The action for such a scalar is
\begin{equation}
\label{action}
S = {1 \over 2\kappa^2}\int d^{10}x{\sqrt g}[{1 \over 2}g^{MN}\partial_M\phi\partial_N\phi] = {V \over 4\kappa^2}\int d^4x\int^{\rho} dr r^5[(\partial_r\phi)^2+h(r)\eta^{\mu \nu}\partial_{\mu}\phi\partial_{\nu}\phi],
\end{equation}
where $V$ is the volume of the $X^5$ and $\rho$ is a UV cutoff to be taken to $\infty$ in the end. We have tacitly switched to Euclidean signature. The indices $M,N$ run over the entire 10-dimensional space, the indices $\mu,\nu$ over 4-dimensional Euclidean space. The equation of motion resulting from this action is
\begin{equation}
\label{coordeq}
(r^{-5}\partial_rr^5\partial_r+h(r)\square)\phi=0.
\end{equation}
Integrating by parts in the action (\ref{action}), we get
\begin{eqnarray*}
\label{fluxes}
S = {V \over 4\kappa^2}\int d^4x\int^{\rho} drr^5[-\phi(r^{-5}\partial_r r^5\partial_r+h(r)\eta^{\mu \nu}\partial_{\mu}\partial_{\nu})\phi+r^{-5}\partial_r(\phi r^5\partial_r\phi)]= \\
=-{V \over 4\kappa^2}[{\cal F}(r)_{r=\rho}-{\cal F}(r)_{r=0}],
\end{eqnarray*}
where ${\cal F}(r) = \phi(r)r^5\partial_r\phi(r)$ is the flux factor. We have used the equation of motion and the fact that there are no boundary terms from integrating by parts in the $x^{\mu}$ directions since the fields are assumed to vanish at 4-dimensional infinity. Going to momentum space, we find
\begin{equation}
\label{momaction}
S = {V \over 4\kappa^2}\int d^4kd^4q\phi_k\phi_q(2\pi)^4\delta^{(4)}(k+q){\cal F}_k,
\end{equation}
where $\phi(x) = \int d^4k\phi_ke^{ikx}$ and 
\begin{equation}
\label{momflux}
{\cal F}_k = [{\tilde \phi}_kr^5\partial_r{\tilde \phi}_k]_0^{\rho}.
\end{equation}
${\tilde \phi}_k$ are momentum modes normalized to ${\tilde \phi}_k(\rho)=1$. From (\ref{corr}), the corresponding 2 point function in momentum space is then
\begin{equation}
\label{2point}
\langle {\cal O}_4(k){\cal O}_4(q)\rangle = {\partial^2S \over \partial\phi_k\partial\phi_q} = (2\pi)^4\delta^{(4)}(k+q){V \over 4\kappa^2}{\cal F}_k.
\end{equation}
Thus, to extract the 2 point function we need to solve the equations of motion for the momentum $k$ Fourier mode of the field $\phi$ with the boundary conditions $\phi(\rho)=1$, $\phi(r \rightarrow 0)$ regular, and find the flux factor ${\cal F}_k$. Note that ultimately, we are interested in terms nonanalytic in $k$, since the analytic terms correspond to contact terms in position space. However, there is a subtlety due to the fact that some of these contact terms diverge as $\r \rightarrow \infty$, and need to be canceled by covariant counterterms. These covariant counterterms will generally change the prefactors in front of the 2-point functions. In the particular case of the minimal massless scalar, though, we will show that the prefactors are unchanged for both AdS and KT backgrounds\footnote{For a discussion of the AdS case see \cite{kw2,kostas}.}.

In the standard AdS/CFT correspondence $h(r)=R^4/r^4$ and the equation of motion (\ref {coordeq}) in momentum space becomes
$$(r^{-5}\partial_rr^5\partial_r-k^2{R^4 \over r^4})\phi =0.$$
Changing variables to $y = kR^2/r$ this is:
\begin{equation}
\label{besseleq}
(y^3\partial_yy^{-3}\partial_y-1)\phi(y)=0.
\end{equation}
This is equivalent to a Bessel equation whose solution with the desired boundary conditions is
$$\phi(y) = {y^2 K_2(y) \over k^2\varepsilon^2K_2(k\varepsilon)}$$
where $\varepsilon = R^2/\rho$ is a UV cutoff. This function has the small $y$ expansion
\begin{equation}
\label{besselexpand}
\phi(y) = 1 -{1 \over 4}y^2 - {1 \over 16}y^4\log y + \ldots
\end{equation}
The logarithmic term gives the leading nonanalytic contribution, so that
$$\langle{\cal O}_4(k){\cal O}_4(-k)\rangle \sim (k\varepsilon)^4 \log (k\varepsilon),$$
or
$$\langle{\cal O}(x_1){\cal O}_4(x_2)\rangle \sim {1 \over |x_1-x_2|^8}.$$

Turning now to the Klebanov-Tseytlin background, our strategy will be simply to repeat the above steps. Consider again the minimal massless scalar. Starting from the action (\ref{action}) with $X^5 = T^{1,1}$, we arrive in the same way as before at the result (\ref{2point}). With the warp factor $h(r)$ defined as in (\ref{ktsol}), the mode $\phi_k(r)$ now satisfies the equation    
\begin{equation}
\label{momeq}
[r^{-5}\partial_r r^5\partial_r -A^2k^2{r_s^4 \over r^4}\log (r/r_s)]\phi(r)=0,
\end{equation}
where we have defined
\begin{equation}
\label{scaledefs}
r_s = r_0e^{-1/4-R^4/2L^4}, \; A^2 = {2L^4 \over r_s^4}.
\end{equation}
Changing variables to
\begin{equation}
\label{ydefs}
y = {Akr_s^2 \over r}, \; Y=Akr_s,
\end{equation}
eq. (\ref{momeq}) becomes
\begin{equation}
\label{yeq}
[y^3\partial_yy^{-3}\partial_y-\log{Y \over y}]\phi(y)=0.
\end{equation}

To find the 2-point function (\ref{2point}), we need to solve eq. (\ref{yeq}) with appropriate boundary conditions. This equation is valid for $y \ll Y$. As $y \rightarrow Y$, we run into a singularity. Recall that we would like to impose the boundary condition that $\phi(y)$ is regular in the IR, i.e. for large $y$. The rigorous way of doing this would be to look at the full Klebanov-Strassler spacetime \cite{ks} which resolves the KT singularity. But the KS solution is rather complicated and one would have no hope of solving the equations analytically. Instead, note that for large enough $k$, i.e. at high energies, $Y$ is a large number. Thus there is a region where $0 \ll y \ll Y$. If we can solve eq. (\ref{yeq}) in this region, we can impose the boundary condition that $\phi$ be regular at large $y$. If $Y \gg 1$ then this boundary condition will mimic the correct one, whatever the details of singularity resolution are. Next, note that if we take $1/Y \ll y \ll Y$, then $|\log y| \ll |\log Y|$ and eq. (\ref{yeq}) reduces to   
\begin{equation}
\label{phiireq}
(y^3\partial_yy^{-3}\partial_y-\log Y)\phi=0.
\end{equation}  
This is Bessel's equation, just like (\ref{besseleq}). Now we take the solution that is regular at large $y$. This is the same solution as we needed in (\ref{besselexpand}):
\begin{equation}
\label{phiir}
\phi_{IR} = B(1 - {1 \over 4}y^2\log Y-{1 \over 16}y^4\log^2 Y\log(\sqrt {\log Y}y)+\ldots),
\end{equation}
where B is an arbitraty constant.
In the UV, i.e. for sufficiently small $y$, we solve (\ref{yeq}) by expanding in $y$, and treating the $\log(Y/y)$ term as a perturbation. Namely, we make the ansatz
\begin{equation}
\label{uvexpand}
\phi = \phi_0 +\phi_1 + \phi_2+ \ldots,
\end{equation}
where 
\begin{equation}
\label{iteration}
[y^3\partial_yy^{-3}\partial_y]\phi_{n+1}=[\log(Y/y)]\phi_n, \; \phi_{-1}=0.
\end{equation}
As before, we impose the boundary condition $\phi(0)=1$, where we have already taken the UV cutoff to infinity. We find
\begin{equation}
\label{phiuv}
\phi_{UV}  = \lgroup 1 - {1 \over 4}y^2\log{Y \over y} + y^4[{1 \over 48}\log^3{Y \over y}+{1 \over 64}\log^2{Y \over y}+{1 \over 128}\log {Y \over y}+C_k]+ \ldots \rgroup
\end{equation}
where $C_k$ is an undetermined constant. The information about the 2 point function is hidden in the constant $C_k$ since all other parts of the above expression are analytic in $k$ (note that $Y/y$ doesn't depend on $k$). 
We will now match $\phi_{UV}$ to $\phi_{IR}$. Let us first identify the overlap region. We said before that the solution (\ref{phiuv}) is valid for small $y$. By looking at this solution we see that it has the form of an expansion in $y^2\log(Y/y)$, so we are allowed to use this solution when $y^2\log(Y/y) \ll 1$. On the other hand, the condition for the validity of eq. (\ref{phiireq}) is $1/Y \ll y \ll Y$. We see that when $Y$ is large, these conditions are compatible and there is an overlap region $1/Y \ll y \ll 1/\sqrt{\log Y}$. In this region we can drop the $\log y$ terms in (\ref{phiuv}) since $|\log y| \ll |\log Y|$. Matching (\ref{phiuv}) to (\ref{phiir}) order by order, the first two terms match if we set $B=1$. However, if we look at the terms multiplying $y^4$, we see that $\phi_{UV}$ has a $\log^3 Y$ term, whereas the leading term in $\phi_{IR}$ is a $\log^2 Y\log\log Y$ term. We must now use the undetermined constant $C_k$ to cancel this leading $\log^3 Y$ term. Thus, we find
\begin{equation}
\label{const}
C_k = -{1 \over 48}\log^3 Y + \ldots=-{1 \over 48}\log^3 Akr_s + \ldots
\end{equation}
where we have kept only the leading nonanalytic term. Using equations (\ref{phiuv},\ref{const},\ref{momflux},\ref{2point}), we are now ready to compute the 2 point function. It is of the form
\begin{equation}
\label{mom2point}
\langle {\cal O}_4(k){\cal O}_4(-k)\rangle \sim {A^4k^4r_s^8 \over \kappa^2} \log^3 Akr_s.
\end{equation} 
After Fourier transforming, this produces a position space 2-point function that behaves as follows:
\begin{equation}
\label{pos2point}
\langle{\cal O}_4(x_1){\cal O}_4(x_2)\rangle \sim g_s^2M^4{\log^2\lgroup r_s^2|x_1-x_2|^2/(g_sM\alpha^{\prime})^2\rgroup \over |x_1-x_2|^8}.
\end{equation}
The range of validity of this result is $Y \gg 1$, which using eqs.(\ref{scaledefs},\ref{scales}) translates into $k \gg r_s/(g_sM\alpha^{\prime})$. The new scale $\Lambda \sim r_s/(g_sM\alpha^{\prime})$ is the only scale that appears in the field theory correlation functions; this is the confinement scale. Our result for the 2-point function is valid at energies higher than this scale, i.e. in the deconfined phase.

Equations (\ref{mom2point},\ref{pos2point}) are our first encounter with powers of the $\log(x)$ appearing in the numerator of the 2-point correlation function. The above derivation shows how the $\log(r)$ factor in the KT metric warp function translates into position (or momentum) space logarithms in the 4-dimensional field theory. These logarithms, with varying powers, will appear in all the correlation functions that we compute. We will discuss their interpretation in the concluding section.

Let us briefly address the issue of renormalization. In terms of the variable $z=Ar_s^2/r$, the on-shell action (\ref{momaction}) needed to compute the 2-point function is proportional to
\beq
\label{onshell}
S \sim \phi(z)z^{-3}\p_z\phi(z),
\eeq
where the action is to be evaluated on the surface $z=\varepsilon$. Since $\phi(z)=\phi_0(1-{1\over 4}k^2z^2\log z+C_kk^4z^4+\ldots)$, where $\phi_0$ is determined by the boundary condition at $z=0$, all terms in this action, starting from the leading divergent contact term $\phi_0^2k^2/{\varepsilon}^2$, will be proportional to $k^2$. To cancel them, we need to introduce covariant counterterms that are local on the surface $z=\varepsilon$ and use the induced metric on that surface. The most divergent such counterterm involving two derivatives (i.e. a power of $k^2$) is $\sim k^2\phi(z)^2/(z^2\log z)$. We see that the leading contribution to this counterterm from the term $C_kk^4z^4$ in the expansion of $\phi(z)$ is at order $z^2 \rightarrow 0$, so $C_k$ does not get renormalized. Thus, in the case of the minimal massless scalar, holographic renormalization does not change the leading order behavior of the 2-point function, including the overall coefficient (which we do not explicitly derive here).     

It is straightforward in principle to extend our method to modes other than the minimal massless scalar. Suppose first that we have succeded in isolating a single mode $\phi$ whose flucutation equation decouples from other modes. Then it is still true that after we perform the change of variables (\ref{ydefs}), in the region $1/Y \ll y \ll Y$ we can replace all logarithms by constants, obtaining exactly solvable equations. We will then choose the solutions of these equations that are regular at large $y$. At small $y$ we can solve the equations by the same sort of iterative expansion as in (\ref{uvexpand},\ref{iteration}) with appropriate boundary conditions. Again, we will find that all the terms in the expansion are analytic in $k$ with the exception of an undetermined constant $C_k$. Matching the UV and IR solutions, we will find as a rule that to match the behavior of the IR solution, we will have to choose $C_k$ so as to cancel the leading log of the UV solution. From this we can then extract the 2-point functions.

In some cases we will encounter the following situation: we are interested in obtaining the correlator for an operator ${\cal O}$ whose dual field $\phi$ couples to other SUGRA fluctuations collectively denoted by $\varphi_i$. Our task is then to solve the fluctuation equations for $\phi,\varphi_i$ subject to the boundary condition
\beq
\label{nondiag}
\phi(x,r \rightarrow \infty) \sim {\hat \phi(x)}, \; \varphi_i(x,r \rightarrow \infty) \rightarrow 0,
\eeq
where ${\hat \phi}$ denotes the boundary condition for the field $\phi$. We will find that we can still perform the iterative expansion (\ref{uvexpand},\ref{iteration}) and solve in the UV for the fields $\phi,\varphi_i$ with boundary conditions (\ref{nondiag}), but that the solution of the equations in the IR limit becomes too cumbersome. Nevertheless, emboldened by our experience with the diagonal modes, we will assume that as before, the arbitrary constant that will appear on our UV expansion must be chosen so as to cancel the leading log coefficient in the critical term when the IR limit is taken. This is an extremely plausible assumption that yields sensible results for the correlation functions; unfortunately, in such cases we are only able to compute the leading order correlator up to a numerical factor\footnote{Note that our inability to determine this numerical factor is {\it not} related to renormalization, but to the difficulty of solving the IR equations and extracting the undetermined constant $C_k$. Given $C_k$, it is straightforward to introduce covariant counterterms and compute its renormalization in the UV.}. 

\section{The R-current and its dual vector}

The cascading $SU(N+M) \times SU(N)$ gauge theory has a classical $U(1)$ R-symmetry that gets broken down to $Z_{2M}$ at the quantum level. As pointed out in \cite{ouyang}, this quantum phenomenon of the gauge theory can be described classically in the supergravity dual. In our SUGRA solution, the R-symmetry corresponds to translation of the angular coordinate $\psi$. Naively, the solution (\ref{ktsol}) is invariant under this gauge symmetry. However, this is not exactly true, because of a subtlety involving the RR 3-form field strength $F_3$. The 3-form given in (\ref{ktsol}) comes from a 2-form potential
\begin{equation}
\label{rrpot}
C_2 = {1 \over 2}M\alpha^{\prime}\psi\omega_2.
\end{equation}
$\psi$ is periodic with period $4\pi$, so this $C_2$ is not single-valued as a function of $\psi$; but it is single-valued up to a gauge transformation. Under a translation $\psi \rightarrow \psi + \varepsilon$,
\begin{equation}
\label{c2trans}
C_2 \rightarrow C_2 + {1 \over 2}M\alpha^{\prime}\varepsilon\omega_2.
\end{equation}
As discussed in \cite{ouyang},  a gauge transformation can only shift $C_2$ by an integer multiple of $\pi\alpha^{\P}\omega_2$, so $\psi \rightarrow \psi+\varepsilon$ is a symmetry if $\varepsilon$ is an integer multiple of $2\pi /M$. Since $\varepsilon$ is defined $mod$ $4\pi$, a $Z_{2M}$ subgroup of $U(1)$ remains a symmetry of the solution.
As usual, the global R-symmetry of the gauge theory becomes gauged in supergravity. As described in \cite{ouyang} this gauge symmetry is spontaneously broken via a Higgs mechanism, and the vector field dual to the gauge theory R-current acquires a mass\footnote{For a supergravity description of spontaneous R-symmetry breaking and R-current correlators in the Coulomb branch of N=4 SYM see \cite{bs}.}. In this section we will derive the equation of motion for this vector and compute its mass. We will then use the method of the previous section to compute the 2-point correlation function of the R-currents. The most general form of this correlation function allowed by the symmetries is
\beq
\label{expectedjrjr}
\langle J^R_{\m}(k)J^R_{\n}(-k)\rangle=A(k^2)\pi_{\m\n}(k)+B(k^2){k_{\m}k_{\n} \over k^2},
\eeq
where
\beq
\label{transproj}
\pi_{\m\n}(k) = \d_{\m\n}-{k_{\m}k_{\n} \over k^2}
\eeq
is the transverse projector in 4 dimensions. A and B are the form factors we would like to compute. Note that if R symmetry is conserved, $\p\c J^R=0$, so $B=0$ in this case. Thus a nonzero $B$ indicates R-symmetry breaking.

The natural metric ansatz for fluctuations of the gauge field is
\begin{equation}
\label{goldmetric}
ds^2 = h^{-1/2}(r)dx_{\mu}dx^{\mu}+h^{1/2}(r)\lgroup dr^2+r^{2}[{1 \over 9}\chi^2+{1 \over 6}\sum_{i=1}^4(g^i)^2]\rgroup,
\end{equation}
where, following \cite{ouyang}, we have defined the 1-form
\begin{equation}
\label{chidef}
\chi = g^5 - 2A_{i}dx^{i},
\end{equation}
which is invariant under the gauge transformations
\begin{equation}
\label{goldgauge}
\psi \rightarrow \psi+2\lambda,  \; A \rightarrow A+d\lambda.
\end{equation}
The RR 3-form field strength varies as:
\begin{equation}
\label{3formvar}
F_3 = {2L^2 \over 9g_s}(g^5+2\partial_{i}\theta dx^{i})\wedge\omega_2={2L^2 \over 9}(\chi+2W_{i}dx^{i})\wedge\omega_2,
\end{equation}
where 
\begin{equation}
\label{wdef}
W_{i} = A_{i}+\partial_{i}\theta
\end{equation}
is a gauge-invariant vector field. In the above formulae the index $i$ ranges over the 5 dimensions $(x^{\m},r)$.
To obtain self-consistent equations of motion, we must also vary the RR scalar $C$, and the RR 4-form $C_4$. The most general variation of the RR 4-form $C_4$ consistent with the symmetries of the problem is 
\begin{equation}
\label{c4var}
\delta C_4=K^0 g^1\wedge\ldots\wedge g^4 + K^1\wedge g^5\wedge dg^5 + K^2\wedge dg^5 +K^3\wedge g^5,
\end{equation}
where the $K^r$s are r-forms. In what follows we will be considering the linearized equations of motion for the fluctuations $W_i$, $\theta$, $K^r_{i_1\ldots i_r}$ around the KT background. The relevant equations of motion are the self-duality condition for ${\tilde F_5}$, the Einstein equations, and the equations of motion for the RR scalar $C$ and the RR 2-form $C_2$:
\begin{equation}
\label{goldeqs1}
\delta\tilde{F_5} = \delta\star{\tilde F_5},
\end{equation}
\begin{equation}
\label{goldeqs2}
\delta R_{i\chi} = \delta\lgroup{g_s\over 4}F_{iPQ}F_{\chi}^{\; PQ}-{g_s \over 4}CH_{iPQ}F_{\chi}^{\; PQ} + {g_s^2 \over 96}{\tilde F_{iPQRS}}{\tilde F_{\chi}^{\; PQRS}}\rgroup,
\end{equation}
\beq
\label{goldeqs3}
\d(d\star dC)=-g_sH_3\wedge\d\star(F_3-CH_3),
\eeq
\beq
\label{goldeqs4}
\d d\star(F_3-CH_3)=g_s\d F_5\wedge H_3.
\eeq
Let us turn to the self-duality equation first. We define the following forms:
\begin{eqnarray}
\label{xforms}
d^0x^{\mu\nu\rho\sigma} = \eta^{\mu\mu^{\prime}}\eta^{\nu\nu^{\prime}}\eta^{\rho\rho^{\prime}}\eta^{\sigma\sigma^{\prime}}\epsilon_{\mu^{\prime}\nu^{\prime}\rho^{\prime}\sigma^{\prime}}= -1, \; dx^{\mu\nu\rho} = \eta^{\mu\mu^{\prime}}\eta^{\nu\nu^{\prime}}\eta^{\rho\rho^{\prime}}\epsilon_{\mu^{\prime}\nu^{\prime}\rho^{\prime}\sigma}dx^{\sigma}, \nonumber \\
d^2x^{\mu\nu} = {1 \over 2}\eta^{\mu\mu^{\prime}}\eta^{\nu\nu^{\prime}}\epsilon_{\mu^{\prime}\nu^{\prime}\rho\sigma}dx^{\rho}\wedge dx^{\sigma}, \; d^3x^{\mu} = {1 \over 6}\eta^{\mu\mu^{\prime}}\epsilon_{\mu^{\prime}\nu\rho\sigma}dx^{\nu}\wedge dx^{\rho}\wedge dx^{\sigma}, \nonumber \\
d^4x = {1 \over 24}\epsilon_{\mu\nu\rho\sigma}dx^{\mu}\wedge dx^{\nu}\wedge dx^{\rho}\wedge dx^{\sigma},
\end{eqnarray}
where $\epsilon_{\mu\nu\rho\sigma}$ is the totally antisymmetric tensor in 4 dimnsions, and $\eta_{\mu\nu}$ is the flat Minkowski metric. The following identities are helpful:
\begin{eqnarray}
\label{goldidentities}
dg^5 \wedge dg^5 = -2g^1\wedge \ldots  \wedge g^4, \nonumber \\
\star(dr\wedge g^1 \wedge \ldots \wedge g^4) = -{12 \over r^3h^2}d^4x\wedge g^5, \nonumber \\
\star(dx^{\mu}\wedge g^1 \wedge \ldots \wedge g^4) = -{12 \over r^3h}d^3x^{\mu}\wedge dr\wedge g^5, \nonumber \\
\star(dx^{\mu}\wedge dr\wedge g^5\wedge dg^5) = {3 \over rh}d^3x^{\mu}\wedge dg^5, \nonumber \\
\star(dx^{\mu}\wedge dx^{\nu}\wedge g^5\wedge dg^5) = -{3 \over r}d^2x^{\mu\nu}\wedge dr\wedge dg^5.
\end{eqnarray}
With $\delta C_4$ given by (\ref{c4var}), and using the identities (\ref{goldidentities}), the variation $\delta {\tilde F_5}$ of the RR 5-form field strength is
\begin{equation}
\label{deltaf5}
\delta\tilde {F_5} = d\delta C_4 + B_2 \wedge \delta F_3 = \nonumber \\
(dK^0+2K^1)\wedge g^1\wedge \ldots \wedge g^4+dK^1\wedge g^5\wedge dg^5+(dK^2-K^3)\wedge dg^5+dK^3 \wedge g^5.
\end{equation}
The variation of its dual is
$$
\delta(\star{\tilde F_5}) = \star(d\delta C_4+B_2\wedge \delta F_3)+(\delta\star){\tilde F_5}= 
$$
$$=\lgroup-{12 \over hr^3}(dK^0+2K^1)_{\m}-{8 \over 9}{R^4+2L^4\log(r/r_0) \over h(r)r^3}W_{\mu}\rgroup d^3x^{\mu}\wedge dr \wedge g^5+
$$
$$+\lgroup-{12 \over h^2r^3}(dK^0+2K^1)_r-{8 \over 9}{R^4+2L^4\log(r/r_0) \over h^2r^3}W_r\rgroup d^4x\wedge g^5+
$$
$$
+{3 \over hr}(dK^1)_{\mu r}d^3x^{\mu}\wedge dg^5-{3 \over r}(dK^1)_{\mu\nu}d^2x^{\mu\nu}\wedge dr \wedge dg^5+
$$
$$+{r \over 3}(dK^2-K^3)_{\m\n r}d^2x^{\m\n}\wedge g^5\wedge dg^5 +{rh \over 3}(dK^2-K^3)_{\m\n\r}dx^{\m\n\r}\wedge dr\wedge g^5\wedge dg^5-
$$
$$-\lgroup {hr^3 \over 12}(dK^3)_{\mu\nu\rho r}dx^{\mu\nu\rho}+{2 \over 27}(R^4+2L^4\log(r/r_0))W_{\mu}dx^{\mu}\rgroup\wedge g^1\wedge \ldots \wedge g^4 +
$$
\begin{equation}
\label{deltastarf5}
+\lgroup{h^2r^3 \over 12}(dK^3)_{\mu\nu\rho\sigma}d^0x^{\mu\nu\rho\sigma}+{2h \over 27}(R^4+2L^4\log(r/r_0))W_r\rgroup dr\wedge g^1 \wedge\ldots\wedge g^4,
\end{equation}
where we have used the identities (\ref{goldidentities}), and set the gauge $\theta=0$. The terms involving $W_i$ come form the variation $\delta\star$ of the Hodge dual, which depends on the metric. At this point, it is convenient to introduce a slightly unusual version of the 5-dimensional Hodge dual, $\star_5$. In this Hodge dual, 4-dimensional indices are raised with flat Minkowski metric, while the $r$-index is raised with $h^{-1}(r)$. Thus for example
\begin{equation}
\label{star5}
\star_5dx_{\mu} = \eta_{\mu\mu^{\prime}}\epsilon^{\mu^{\prime}}_{\nu\rho\sigma}dx^{\nu}\wedge dx^{\rho}\wedge dx^{\sigma}\wedge dr, \; \star_5dr = h^{-1}(r)\epsilon_{\mu\nu\rho\sigma}dx^{\mu}\wedge dx^{\nu}\wedge dx^{\rho}\wedge dx^{\sigma},
\end{equation}
etc. With this notation, we find that the self-duality condition (38) reduces to the following two equations:
\begin{equation}
\label{expself1}
K^3 = dK^2 +{3 \over r}\star_5 dK^1,
\end{equation}
\begin{equation}
\label{expself2}
dK^3+{12 \over hr^3}\star_5(dK^0+2K^1)+{8(R^4+2L^4\log{r \over r_0}) \over 9hr^3}\star_5W=0.
\end{equation}
Without loss of generality, we can set $K^0=0$ and $K^2=0$. Then the above reduces to
\begin{equation}
\label{keq1}
K^3 = {3 \over r}\star_5 dK,
\end{equation}
\begin{equation}
\label{keq2}
hr\partial_r{1 \over hr}(\partial_rK_{\m}-\p_{\m}K_r)+h\p_{\n}(\p_{\n}K_{\m}-\p_{\m}K_{\n})-{8 \over r^2}K_{\mu}-{8(R^4+2L^4\log(r/r_0)) \over 27r^2}W_{\m}=0,
\end{equation}
\beq
\label{keq3}
h\lgroup\square K_r-\p_r(\p\cdot K)\rgroup-{8 \over r^2}K_{r}-{8(R^4+2L^4\log(r/r_0)) \over 27r^2}W_{r}=0,
\eeq
where we now denote the vector $K^1$ simply by $K$. We have separated the $x^{\m}$ and $r$ components of the equations of motion for $K$; $\p\cdot K$ denotes the 4-dimensional divergence.

We now turn to the Einstein equations (\ref{goldeqs2}). The variation of the Ricci tensor in terms of the metric variation is
\begin{equation}
\label{riccivar}
\delta R_{MN} = \square_{10} h_{MN}+D_MD_Nh_P^{\; P}-D_MD^Ph_{PN}-D_ND^Ph_{MP}-2R_{MPSN}h^{PS}+R_M^{\; P}h_{PN}+R_N^{\; P}h_{MP},
\end{equation}
where $\square_{10}$ is the 10-dimensional Laplace operator, $h_{MN}=\delta g_{MN}$, and all covariant derivatives, as well as raised indices, are taken with respect to the background metric. Plugging this into eq. (\ref{goldeqs2}) and using equations (\ref{goldmetric},\ref{3formvar}) as well as the equations of motion (\ref{keq1},\ref{keq2},\ref{keq3}) for $K^1$ and $K^3$, this becomes
\beq
\label{weq1}
{1 \over hr^7}\partial_r hr^7 (\partial_rW_{\m}-\p_{\m}W_r) + h\p_{\n}(\p_{\n}W_{\mu}-\p_{\m}W_{\n})-{8L^4 \over hr^6}W_{\mu}- \nonumber \\ 
-16{(R^4+2L^4\log(r/r_0))^2 \over h^2r^{10}}(W_{\mu}+{27 \over R^4+2L^4\log (r/r_0)}K_{\mu})=0,
\eeq
\beq
\label{weq2}
h\lgroup\square W_r-\p_r(\p\cdot W)\rgroup-{8L^4 \over hr^6}(W_{r}-{3C \over 2r})-
 \nonumber \\ 
-16{(R^4+2L^4\log(r/r_0))^2 \over h^2r^{10}}(W_r+{27 \over R^4+2L^4\log (r/r_0)}K_r)=0,
\eeq
where again we have separated the $x^{\m}$ and $r$ components.

In general, whenever we have Lorentz-invariant equations of motion involving a vector mode $A_{\m}$\footnote{By ``vectors'' we mean vectors with respect to the 4-dimensional Lorentz group; the $r$-components $A_r$ are scalars with respect to that group.}, they can be separated into transverse and longitudinal components ${\t A_{\m}}$ and $\p\cdot A$ by setting
\beq
\label{sep}
A_{\m} = {\t A_{\m}} + {\p_{\m}(\p\c A)\over \square}.
\eeq
The transverse mode ${\t A_{\m}}$ that satisfies $\p\c{\t A}=0$ then decouples from all scalar fluctuations and can only couple to other transverse vectors.

In the present case we have two vector fields $K_{\m}$, $W_{\m}$. Defining their transverse parts as above, we find that they only couple to each other, and satisfy the equations: 
\beq
\label{transeq1}
({1 \over hr^7}\partial_r hr^7 \partial_r + h\square){\t W}_{\mu}-{8L^4 \over hr^6}{\t W}_{\mu}- \nonumber \\
-16{(R^4+2L^4\log(r/r_0))^2 \over h^2r^{10}}({\t W}_{\mu}+{27 \over R^4+2L^4\log (r/r_0)}{\t K}_{\mu})=0,
\eeq
\begin{equation}
\label{transeqeq2}
(hr\partial_r{1 \over hr}\partial_r+h\square){\t K}_{\m}-{8 \over r^2}{\t K}_{\mu}-{8(R^4+2L^4\log(r/r_0)) \over 27r^2}{\t W}_{\m}=0.
\end{equation}
These coupled equations can be diagonalized by taking the following linear combinations:
\begin{equation}
\label{diag}
 W^1 =  W - {54 \over hr^4}K, \; W^2 = W + {27 \over hr^4}K.
\end{equation}
Then the transverse components ${\t W}^1$ and ${\t W}^2$ satisfy the equations:
\begin{equation}
\label{w1eq}
\lgroup{1 \over hr^7}\partial_r hr^7 \partial_r + h\square-{4L^8 \over r^2(R^4+2L^4(\log(r/r_0)+{1 \over 4}))^2}\rgroup {\t W}^1_{\mu}=0,
\end{equation}
\beq
\label{w2eq}
\lgroup{1 \over hr^7}\partial_r hr^7 \partial_r + h\square- \nonumber \\
-4{6R^8+3R^4L^4+(24R^4L^4+6L^8)\log(r/r_0)+24L^8\log^2(r/r_0) \over r^2(R^4+2L^4(\log(r/r_0)+{1 \over 4}))^2}\rgroup {\t W}^2_{\mu}=0.
\eeq
By inspection we see that $W^2$ is massive in the $AdS_5 \times T^{1,1}$ limit where we take $L=0$; the mode we're interested in is $W^1$. This is the Goldstone vector that acquires a mass, corresponding to the spontaneous breaking of R-symmetry. 

The authors of ref. \cite{bdfp} made a general prediction for the mass of a vector associated with such symmetry breaking. We can now compare that prediction to our result. Eq. (193) of \cite{bdfp} reads
\begin{equation}
\label{dewolfe}
(e^{-2T}\partial_qe^{2T}\partial_q+e^{-2T}\square+2{\partial^2T \over \partial q^2}){\t V}_{\mu}=0,
\end{equation}
where $V_{\mu}$ is related to $W^1_{\mu}$ by a rescaling and $q,T$ are such that the reduced metric in 5 dimensions is
\begin{equation}
\label{reducedmetricform}
ds_5^2  = dq^2 + e^{2T}dx_{\mu}dx^{\mu}.
\end{equation}
In terms of the $r$ coordinate, the reduced KT metric (6) in 5 dimensions is
\begin{equation}
\label{ktreduced}
ds_5^2 = (h(r)r^4/R_0^4)^{5/6}(h^{1/2}(r)dr^2+h^{-1/2}(r)dx_{\mu}dx^{\mu}),
\end{equation}
where $R_0$ is some reduction scale. Comparing this to (\ref{reducedmetricform}) we get
\beq
\label{dw1}
dq = {dr \over rR_0^{5/3}}(R^4+2L^4(\log(r/r_0)+1/4))^{2/3},
\eeq
\beq
\label{dw2}
e^{2T} = r^2R_0^{-10/3}(R^4+2L^4(\log(r/r_0)+1/4))^{1/3}.
\eeq
We now transform (\ref{dewolfe}) to the $r$ coordinate. To obtain agreement between the kinetic terms we also need to rescale:
\begin{equation}
\label{dewolferescale}
V_{\mu} = (h(r)r^4/R_0^4)^{2/3}W^1_{\mu}.
\end{equation}
Plugging the above expressions in, we get
\begin{equation}
\label{w1eqagain}
\lgroup{1 \over hr^7}\partial_r hr^7 \partial_r + h\square-{4L^8 \over r^2(R^4+2L^4(\log(r/r_0)+{1 \over 4}))^2}\rgroup {\t W}^1_{\mu}=0,
\end{equation}
which is precisely our eq.(\ref{w1eq}).
In terms of the gauged supergravity conventions of ref. \cite{bdfp}, the vector $W^1$ has picked up a mass
\begin{equation}
\label{mass}
m^2 = -2{d^2T \over dq^2} = {4 \over \alpha^{\prime}(3\pi)^{3/2}}{(g_sM)^2 \over (g_sN)^{3/2}},
\end{equation}
where we have used the relations (\ref{scales}).

Before proceeding to derive the remaining equations of motion for the scalar sector, let us first calculate the leading order 2-point functions $\langle {\t J}^{1,2}_{\mu}(x){\t J}^{1,2}_{\nu}(x^{\prime})\rangle$ for the transverse components of the gauge theory currents $J^{1,2}$ dual to the supergravity modes $W^{1,2}$ we have found. We follow the method outlined in the previous section. Using the change of variables (18,19), eqs. (52,53) in momentum space become
\begin{equation}
\label{w1yeq}
\lgroup {y \over \log(Y/y)}\partial_yy^{-1}\log(Y/y)\partial_y-{1 \over y^2\log^2(Y/y)}-\log(Y/y)\rgroup {\t W}^1 = 0,
\end{equation}
\begin{equation}
\label{w2yeq}
\lgroup {y \over \log(Y/y)}\partial_yy^{-1}\log(Y/y)\partial_y-{24\log^2(Y/y)-6\log(Y/y)+1 \over y^2\log^2(Y/y)}-\log(Y/y)\rgroup {\t W}^2 = 0.
\end{equation}
In the IR region $1/Y \ll y \ll Y$, these reduce to Bessel equations:
\begin{equation}
\label{w1ireq}
\lgroup y\partial_yy^{-1}\partial_y-\log Y\rgroup {\t W}^1_{IR} = 0,
\end{equation}
\begin{equation}
\label{w2ireq}
\lgroup y\partial_yy^{-1}\partial_y-{24 \over y^2}-\log Y\rgroup {\t W^2}_{IR} = 0,
\end{equation}
where we have also used $Y \gg 1$.  Note that in going from eq. (\ref{w1yeq}) to (\ref{w1ireq}), the mass term $1/(y^2\log^2(Y/y))$ is left out since it becomes suppressed by a factor of $1/(\log Y)^2$. This is the term responsible for the anomalous dimension of the R-current, so we see that this anomalous dimension will not show up in our calculations, which are at leading order in high energy. If this term were included, it would modify the order of the Bessel function, and thus the power of $k$ in the correlator. The solutions of these Bessel equations (\ref{w1ireq},\ref{w2ireq}) that remain regular at large $y$ are, up to a multiplicative constant
\begin{equation}
\label{w1ir}
{\t W}^1_{IR} \sim 1 + B_1y^2\log Y \log\log Y + \ldots
\end{equation}
\begin{equation}
\label{w2ir}
{\t W}^2_{IR} \sim {1 \over y^4\log Y} + \ldots + B_2y^6\log^4(Y)\log\log Y + \ldots
\end{equation}
where we have only included the terms relevant to our matching. $B_1$, $B_2$ are constants whose exact value will not matter to us. In the UV region $y \ll 1$, we perform an iterative expansion similar to eqs. (23,24). We find
\begin{equation}
\label{w1uv}
{\t W}^1_{UV} \sim {1 \over 2}{2\log(Y/y)-1 \over \log(Y/y)}-{1 \over 6}y^2\log^2(Y/y)+C_1{y^2 \over \log(Y/y)}+ \ldots
\end{equation}
\begin{equation}
{\t W}^2_{UV} \sim {1 \over y^4\log(Y/y)} - \ldots -{1 \over 7200}y^6\log^5(Y/y)+C_2y^6 +\ldots
\end{equation}
where again we omitted terms not relevant to the matching. Performing the matching, we find that, just as in the case of the minimal scalar, the UV expansions $W^{1,2}_{UV}$ have a higher power of $\log Y$ in the coefficient of the critical power of $y$ than the IR functions that remain regular at large $y$. We must use the arbitrary constants $C_1$, $C_2$ to cancel this leading log; these constants then encode the leading-order 2-point functions. Thus we have
\begin{equation}
\label{c12}
C_1 = {1 \over 6}\log^3Y+\ldots, \; C_2 = {1 \over 230400}\log^5Y+\ldots
\end{equation}
Restoring the $k$-dependence of $y$ and $Y$, we obtain
\begin{equation}
\label{w1ruv}
{\t W}^1_{UV} \sim 1 + {L^4k^2 \over 3r^2\log(r/r_s)}\log^3(k/\Lambda)+\ldots,
\end{equation}
\begin{equation}
\label{w2ruv}
{\t W}^2_{UV} \sim {r^4 \over \log(r/r_s)}+\ldots+{L^{20}k^{10}\log^5(k/\Lambda) \over 7200r^6}+\ldots
\end{equation}
where $\Lambda$ is given by $\Lambda \sim r_s/(M\alpha^{\prime})$. As usual, the momentum space 2-point functions are proportional to the lowest-order nonanalytic terms in k, so we have
\begin{equation}
\label{j1mom}
\langle {\t J}^R_{\mu}(k){\t J}^R_{\nu}(-k)\rangle \sim g_s^2M^4\pi_{\mu\nu}(k)k^2\log^3(k/\Lambda),
\end{equation}
\begin{equation}
\label{j2mom}
\langle {\t J}^2_{\mu}(K){\t J}^2_{\nu}(-k)\rangle \sim g_s^{10}M^{12}(\alpha^{\P})^8\pi_{\mu\nu}(k)k^{10}\log^5(k/\Lambda),
\end{equation}
where we have renamed $J^1 \equiv J^R$ since it is in fact the R-current. The transverse projector $\pi_{\m\n}$ is defined in (\ref{transproj}). These translate into position space 2-point functions
\begin{equation}
\label{j1pos}
\langle {\t J}^R_{\mu}(x){\t J}^R_{\nu}(x^{\prime})\rangle \sim  g_s^2M^4(\delta_{\mu\nu}-{\p_{\m}\p_{\n} \over \square}){\log^2(\Lambda|x-x^{\prime}|) \over |x-x^{\prime}|^6},
\end{equation}
\begin{equation}
\label{j2pos}
\langle {\t J}^2_{\mu}(x){\t J}^2_{\nu}(x^{\prime})\rangle \sim  g_s^{10}M^{12}(\alpha^{\P})^8(\delta_{\mu\nu}-{\p_{\m}\p_{\n} \over \square}){\log^4(\Lambda|x-x^{\prime}|) \over |x-x^{\prime}|^{14}}.
\end{equation}
We now make a brief digression to compute yet another transverse current-current correlator; the purpose is to once again demontrate the general pattern of these calculations. First, we note\footnote{This observation is due to Igor Klebanov.} that if we vary the NS-NS 2-form as
\begin{equation}
\label{vecmode}
B_2 \rightarrow B_2 + A_idx^i\wedge g^5,
\end{equation}
while leaving all other fields in the KT solution constant, the vector field $A_i$ decouples from all other modes. Inserting the variation (\ref{vecmode}) into the equation of motion (2) for $B_2$, and, as usual, expanding to linear order in $A_i$, we find that the transverse 4-dimensional components ${\t A}_{\mu}$ obey the equations of motion
\begin{equation}
\label{aeq}
\lgroup r^{-3}\partial_rr^3\partial_r+h\square-{8 \over r^2}\rgroup {\t A}_{\mu}=0.
\end{equation}
Performing once again the change of variables (18,19), this becomes
\begin{equation}
\label{ayeq}
\lgroup y\partial_yy^{-1}\partial_y-{8 \over y^2}-\log(Y/y)\rgroup {\t A}_{\mu}=0.
\end{equation}
The IR and UV expansions are obtained in the usual way:
\begin{equation}
\label{air}
{\t A}_{IR} \sim {1 \over y^2}+ \ldots +By^4\log^3Y\log\log Y+\ldots
\end{equation}
\begin{equation}
\label{auv}
{\t A}_{UV} \sim {1\over y^2}- \ldots -{1 \over192}y^4\log^4(Y/y)+Cy^4+\ldots
\end{equation}
Again we see that in the matching region the UV expansion has a higher power of $\log Y$ in the critical term coefficient, and we use the constant $C$ to cancel it. Thus
\begin{equation}
\label{ca}
C = {\log^4Y \over 192}+\ldots
\end{equation}
With the usual transformations, we find a corresponding position space 2-point function
\begin{equation}
\label{apos}
\langle {\t J}^{A}_{\mu}(x){\t J}^{A}_{\nu}(x^{\prime})\rangle \sim (\delta_{\mu\nu}-{\p_{\m}\p_{\n} \over \square}){\log^3(\Lambda|x-x^{\prime}|)\over |x-x^{\prime}|^{10}}.
\end{equation}
The lesson from the above calculations is that, to leading order, the high energy behavior of the 2-point functions can be extracted from the UV iterative expansion alone; the matching with the IR solution always has the consequence that we must choose the undetermined constant $C_k$ in the UV expansion in such a way as to cancel the leading log coefficient of the critical power of $y$ in the IR limit. We will now use this shortcut to compute the longitudinal part of the R-current correlator $\langle J^{R}_{\m}J^R_{\n}\rangle$. First, we need to derive the remaining scalar equations of motion. Eqs. (\ref{goldeqs3},\ref{goldeqs4}) yield
\beq
\label{axioneq}
(r^{-5}\p_rr^5\p_r+h\square)C+{16L^4 \over 3hr^5}(W_r-{3C \over 2r})=0,
\eeq
\beq
\label{f3eq}
(\p\c W)=-{108 \over r^5h^2}\lgroup K_r+{R^4+2L^4\l \over 27}W_r\rgroup-{1 \over 2r}\p_r\lgroup{r \over h}(W_r-{3C \over 2r})\rgroup.
\eeq
Also, by taking divergences of eqs. (\ref{keq2},\ref{weq1}) we find
\beq
\label{divkeq}
hr\partial_r{1 \over hr}\partial_r(\p\c K)-{8 \over r^2}(\p\c K)-{8(R^4+2L^4\log(r/r_0)) \over 27r^2}(\p\c W)-hr\partial_r{1 \over hr}\square K_r=0,
\eeq
\beq
\label{divweq}
{1 \over hr^7}\partial_r hr^7 \partial_r(\p\c W)-{8L^4 \over hr^6}{\p\c W}- \nonumber \\
-16{(R^4+2L^4\log(r/r_0))^2 \over h^2r^{10}}(\p\c W+{27 \over R^4+2L^4\log (r/r_0)}\p\c K)-{1 \over hr^7}\partial_r hr^7\square W_r=0.
\eeq
Equations (\ref{axioneq}--\ref{divweq}), along with eqs. (\ref{keq3},\ref{weq2}) are the equations of motion in the scalar sector. The independent fields may be taken to be the scalars $C, W_r, K_r$; it is possible to check, as must of course be the case, that these six equations for three independent fields are consistent.

We now want to extract the longitudinal part of the $\langle J^R_{\m}J^R_{\n}\rangle$ correlator. As discussed above, we only need to solve the equations to the critical order in the UV expansion. Recall once again that in practice, this means that in the 0-th approximation, we drop all the '$\square$' terms, since they scale as $r^{-4}$, whereas all other terms scale as $r^{-2}$. The '$\square$' terms operating on the 0-th order solution are then included in the equations for the 1-st order solutions, etc. Since the vector dual to the R-current operator is $W^1$ as defined in eq. (\ref{diag}), we impose the boundary conditions
\beq
\label{goldscalarbound}
\p\c W^1(r,x)(r \rightarrow \infty) \rightarrow \p\c W^1(x), \; \; \; \p\c W^2(r,x)(r \rightarrow \infty) \rightarrow 0, \nonumber \\
C(r,x)(r \rightarrow \infty) \rightarrow 0, \; \; W_r(r,x)(r \rightarrow \infty) \rightarrow 0, \; \; K_r(r,x)(r \rightarrow \infty) \rightarrow 0.
\eeq
As usual, we transform to momentum space, and seek normalized solutions of the form
\beq
\label{goldscalarkbound}
\p\c W^1(r,k)(r \rightarrow \infty) \rightarrow 1, \; \; \; \p\c W^2(r,k)(r \rightarrow \infty) \rightarrow 0, \nonumber \\
C(r,k)(r \rightarrow \infty) \rightarrow 0, \; \; W_r(r,k)(r \rightarrow \infty) \rightarrow 0, \; \; K_r(r,k)(r \rightarrow \infty) \rightarrow 0.
\eeq
We are indeed able to find a solution with these boundary conditions. To first order in the UV expansion (i.e. iterating once), the solution is:
\beq
\label{jruvsols}
\p\c W^1(r) = 1 - {1 \over 3}{k^2L^4 \over r^2}\log(r/r_s)+C_1{1 \over r^2\log(r/r_s)}, \; \; \p\c W^2(r) = -{1 \over 72}{k^2L^4 \over r^2}, \nonumber \\
C(r)=-{2 \over 9}{L^4 \over r^2}+{1 \over 27}{k^2L^8 \over r^4}\log^3(r/r_s)-{2 \over 3}C_1{L^4 \over r^4}\log(r/r_s), \nonumber \\
W_r(r)=-{1 \over 3}{L^4 \over r^3}\log(r/r_s)+{2 \over 9}{k^2L^8 \over r^5}\log^3(r/r_s)+2C_1{L^4 \over r^5}\log(r/r_s), \nonumber \\
K_r(r)={2 \over 81}{L^8 \over r^3}\log^2(r/r_s)-{4 \over 243}{k^2L^{12} \over r^5}\log^4(r/r_s)-{4 \over 27}C_1{L^8 \over r^5}\log^2(r/r_s),
\eeq
where we have kept only the leading log terms at each power of $r$, and $C_1$ is the undetermined constant that contains the information we need. Looking at the solution (\ref{jruvsols}), we see that in going to the IR limit $1/Y \ll y \ll Y$ with $y,Y$ defined as in (\ref{scaledefs},\ref{ydefs}), to cancel the leading log in the critical term, we have to choose the undetermined constant
\beq
\label{jrc1}
C_1 \sim k^2L^4 \log^2(k/\Lambda),
\eeq
which leads to the longitudinal momentum space R-current 2-point function
\beq
\label{jrlong}
\langle J^{\parallel}_{\m}(k)J^{\parallel}_{\n}(-k)\rangle \sim g_s^2M^4k_{\m}k_{\n}\log^2(k/\Lambda).
\eeq
Mulitplying the above by $k_{\m}k_{\n}$, we also get
\beq
\label{divjrdivjr}
\langle \p\c J^R(k)\p\c J^R(-k)\rangle \sim g_s^2M^4k^4\log^2(k/\Lambda).
\eeq
Thus, the 2-point function of the R-anomaly scalar $\p\c J^R$ has a different leading-order logarithmic behavior from that of the minimal scalar (\ref{mom2point}) found in the previous section.

Unfortunately, the above analysis does not in itself allow us to determine the numerical value of the constant $C_1$; to do that, we would need to diagonalize the fluctuation equations in the IR and match them in detail to the UV solution (\ref{jruvsols}), which we are unable to do at present. Thus, by combining eqs. (\ref{j1mom},\ref{jrlong}) we can write the total R-current 2-point function as 
\beq
\label{jr2point}
\langle J^R_{\m}(k)J^R_{\n}(-k)\rangle = g_s^2M^4\lgroup C_0\pi_{\mu\nu}(k)k^2\log^3(k/\Lambda)+C_1k_{\m}k_{\n}\log^2(k/\Lambda)\rgroup,
\eeq
where we could in principle compute the prefactor $C_0$ exactly (though we do not bother do that here), but we have not been able to determine the prefactor $C_1$. 

\section{The EM tensor and the graviton}  

In this section we would like to compute the short-distance behavior of the field theory energy-momentum tensor 2-point function $\langle T_{\mu\nu}(x)T_{\rho\sigma}(x^{\prime})\rangle$. This is an object of interest because in a conformal theory, the structure of this correlator is completely determined by conformal symmetry. Thus, any deviation from the CFT result will exhibit some of the structure of the breaking of conformal symmetry, and yield information about the flow of the beta function. More specifically, the most general form of the $\langle TT\rangle$ correlator allowed by translation invariance is
\beq
\label{expectedtt}
\langle T_{\m\n}(k)T_{\r\s}(-k)\rangle =  C(k^2)\pi_{\m\n\r\s}(k)+D(k^2)\pi_{\m\n}(k)\pi_{\r\s}(k),
\eeq
where $\pi_{\m\n}$ is the transverse projector defined in eq. (\ref{transproj}), and
\beq
\label{ttprojector}
\pi_{\m\n\r\s} = {1 \over 2}(\pi_{\m\s}\pi_{\n\r}+\pi_{\m\r}\pi_{\n\s})-{1 \over 3}\pi_{\m\n}\pi_{\r\s}
\eeq
is the transverse traceless projector. Our purpose is to compute the form factors $C$ and $D$. Note that in a scale invariant theory, $T_{\m}^{\; \m}=0$, and therefore $D=0$; thus a nonzero $D$ is an indication of the trace anomaly. Moreover, conformal symmetry dictates $C(k) \sim k^4$, so a nontrivial $C$ also manifests the breaking of conformal symmetry.

The supergravity field dual to the EM-tensor operator $T_{\mu\nu}$ is the graviton $\gamma^{\mu}_{\; \nu}$ along the brane directions, normalized with respect to the background metric (see ref.\cite{liu}). In other words, we vary the 4-dimensional part of the metric as follows:
\begin{equation}
\label{metricvar}
g_{\mu\nu} \rightarrow g_{\mu\nu}+g_{\mu\rho}\gamma^{\rho}_{\; \nu}.
\end{equation}
The 2-point function is then
\begin{equation}
\label{ttform}
\langle T_{\mu\nu}(x)T_{\rho\sigma}(x^{\prime})\rangle = {\delta^2S(\gamma,\phi_i) \over \delta {\hat \g}_{\mu\nu}(x)\delta {\hat \g}_{\rho\sigma}(x^{\prime})},
\end{equation}
where $\gamma$ denotes the graviton along the branes and $\phi_i$ denotes collectively all other supergravity fields, and the action $S$ is evaluated at the solution to the linearized SUGRA equations of motion with the boundary conditions
\begin{equation}
\label{ttboundary}
\gamma^{\mu}_{\; \nu}(x,r \rightarrow \infty) \rightarrow {\hat \g}_{\mu\nu}(x), \; \phi_i(x, r \rightarrow \infty) \rightarrow 0.
\end{equation}
In what follows, we will not be careful about distinguishing upper and lower indices; rather, we will assume that all indices are raised and lowered with flat metric, and tacitly insert appropriate factors of $h(r)$ as needed.

Our task is to solve the linearized SUGRA equations of motion around the KT background with boundary conditions (\ref{ttboundary}). The graviton $\gamma_{\mu\nu}$ couples to other fields, so we must include their fluctuations as well. To simplify the calculations somewhat, we will set $R=0$ in the definition of $h(r)$ (see eq. 6). A self-consistent ansatz is:
\beq
\label{ttansatz}
ds^2 = h^{-1/2}(r)\lgroup\eta_{\mu\nu}+\gamma_{\mu\nu}(x,r)\rgroup dx^{\mu}dx^{\nu}+ \nonumber \\
+h^{1/2}(r)\lgroup\gamma_{rr}(x,r)dr^2+r^2[{1 \over 6}(1+{1 \over 4}s_1(x,r))\sum_{i=1}^4(g^i)^2+{1 \over 9}(1+s_2(x,r))(g^5)^2]\rgroup, \nonumber \\
B_2 = {2L^2 \over 3}\log(r/r_0)(1+\delta B(x,r))\omega_2, \; C_4 = (\partial_rh^{-1})(1+\delta C(x,r))d^4x, \; \Phi = \delta\Phi(x,r),
\eeq   
with all other fields equal to their background values. The self-duality condition (38) for ${\tilde F_5}$ allows us immediately to solve for the field $\delta C$:
\begin{equation}
\label{deltac}
\delta C = \delta B+{1 \over 2}(\gamma+\gamma_{rr}-s),
\end{equation}
where we have defined the traces
\begin{equation}
\label{tracedefs}
\gamma = \d^{\mu\nu}\gamma_{\mu\nu}, \; s=s_1+s_2.
\end{equation}
We now turn our attention to the Einstein equations (1). Using the expansion (49) and looking at the '$\mu\nu$' and '$\mu r$' components, we obtain
\beq
\label{gravmunu}
(h\square+\p_r^2+{5 \over r}\p_r)\g_{\mu\nu}+h\p_{\mu}\p_{\nu}(\g+\g_{rr}+s)-h\p_{\mu}\p_{\rho}\g_{\rho\nu}-h\p_{\nu}\p_{\rho}\g_{\mu\rho}+ \nonumber \\
+\d_{\mu\nu}[{1 \over r}{\l \over \l+1/4}\p_r(\g+s-\g_{rr}-2\dB)- \nonumber \\
-4{32\ls+4\l+1 \over r^2(4\l+1)^2}(\g_{rr}-s+2\dB)]=0
\eeq
and
\begin{equation}
\label{gravmur}
\p_r\p_{\mu}\g_{\mu\nu}+({5 \over r}+{h^{\P} \over 2h})\p_{\mu}\g_{rr}-\p_{\mu}\p_{r}(\g+s)-({1 \over r}+{h^{\P} \over 2h})\p_{\mu}s-{4 \over r}{\l \over \l+1/4}\p_{\mu}\dB=0,
\end{equation}
where we have used the relation (\ref{deltac}). Following ref. \cite{af}, we define the transverse traceless part of the graviton ${\bar \g_{\mu\nu}}$ as
\beq
\label{gammabardef} 
{\bar \g_{\mu\nu}}=\g_{\mu\nu}-{1 \over \square}\p_{\mu}\p_{\rho}\g_{\rho\nu}-{1 \over \square}\p_{\nu}\p_{\rho}\g_{\mu\rho}+{\p_{\mu}\p_{\nu} \over \square^2}\p_{\rho}\p_{\sigma}\g_{\rho\sigma}+ \nonumber \\
+{1 \over 3}({\p_{\mu}\p_{\nu} \over \square}-\d_{\mu\nu})(\g-{1 \over \square}\p_{\rho}\p_{\sigma}\g_{\rho\sigma}).
\eeq
This tensor satisfies:
\begin{equation}
\label{gammabarrelations}
\d^{\m\n}\bar{\g}_{\m\n}=0, \; \p_{\mu}{\bar \g}_{\mu\nu}=0.
\end{equation}
From the index structure of eq. (\ref{gravmunu}), it is clear that ${\bar \g}_{\mu\nu}$ decouples from all other fields and satisfies the equation:
\begin{equation}
\label{gammabareq}
(h\square+r^{-5}\p_rr^5\p_r){\bar \g}_{\mu\nu}=0.
\end{equation}
This is precisely the equation for the minimal massless scalar, so following the steps outlined in section 3, we obtain the transverse traceless (TT) part of the energy-momentum tensor 2-point function:
\beq
\label{TT}
\langle T_{\m\n}(k)T_{\r\s}(-k)\rangle_{TT} \sim g_s^2M^4\pi_{\m\n\r\s}(k)k^4\log^3(k/\Lambda),
\eeq
where the transverse traceless projector $\pi_{\m\n\r\s}$ was defined in eq. (\ref{ttprojector}). In terms of the form factors $C,D$ defined in (\ref{expectedtt}), we find $C(k) \sim k^4\log^3k$. This is different from the $k^4$ behavior required by conformal symmetry.

Of course, eq.(\ref{gammabareq}) does not exhaust the information contained in eq.(\ref{gravmunu}). By substituting (\ref{gammabardef}) into (\ref{gravmunu}) and taking the trace or multiplying it by $\p_{\mu}$, we obtain
\begin{equation}
\label{gammatraceeq}
r^{-5}\p_rr^5\p_r\g+h\lgroup-2\p_{\mu}\p_{\nu}\g_{\mu\nu}+2\square\g+\square(\g_{rr}+s)\rgroup+4V=0,
\end{equation}
\begin{equation}
\label{iphieq}
r^{-5}\p_rr^5\p_r(\p_{\nu}\g_{\mu\nu})+h\lgroup(-\p_{\mu}\p_{\rho}\p_{\sigma}\g_{\r\s}+\p_{\m}\square(\g+\gr+s)\rgroup+\p_{\m}V=0,
\end{equation}
where we have defined
\beq
\label{Vdef}
V = {1 \over r}{\l \over \l+1/4}\p_r(\g+s-\g_{rr}-2\dB)- \nonumber \\
-4{32\ls+4\l+1 \over r^2(4\l+1)^2}(\g_{rr}-s+2\dB).
\eeq
Next, we  define a new scalar field $\f$ by
\beq
\label{phidef}
\p_{\n}\g_{\m\n}=\p_{\m}\f+C_{\m}(x).
\eeq
$\f$ is well-defined because eq. (90) shows that the $r$-derivative of the vector $\p_{\n}\g_{\m\n}$ is a 4-dimensional gradient. Hence it is itself a 4-dimensional gradient up to an $r$-independent vector $C_{\m}(x)$. With the definition (\ref{phidef}), we have
\beq
\label{geq}
r^{-5}\p_rr^5\p_r\g+h\lgroup\square(2\g-2\f+\gr+s)-2\p_{\m}C_{\m}\rgroup+V=0,
\eeq
\beq
\label{phieq}
\p_r\f+X=0,
\eeq 
where we have also defined 
\beq
\label{xdef}
X=({5 \over r}+{h^{\P} \over 2h})\g_{rr}-\p_{r}(\g+s)-({1 \over r}+{h^{\P} \over 2h})s-{4 \over r}{\l \over \l+1/4}\dB.
\eeq
Let us now write out the equations of motion for the other scalar fields $s_1,s_2,\gr,\dB,\dP$. Expanding eqs. (1-3) linearly in these fields with the ansatz (86), we get
\beq
\label{dileq}
(h\square+r^{-5}\p_rr^5\p_r)\dP-{8L^4 \over r^6h}\lgroup r\p_r[\l\dB]+{1 \over 2}(s_2-\gr-2\dP)\rgroup=0,
\eeq
\beq
\label{beq} 
(h\square+{h \over r\ls}\p_r{r\ls \over h}\p_r)\dB+{1 \over 2r\l}\p_r(\g-\gr+s_2-2\dP)+ \nonumber \\
+{h(\p_rh^{-1}) \over 2r\l}(-2\gr+2s_2+s_1-2\dP)=0,
\eeq
\beq
\label{greq}
h\square\gr+\p_r^2(s+\g)+6{\l \over r(\l+1/4)}\p_r\dB+3{\l \over r(\l+1/4)}\p_r\g- \nonumber \\
-{64\ls+36\l+5 \over r(4\l+1)^2}\p_r\gr+{2(8\ls+6\l+1) \over r(4\l+1)^2}\p_rs- \nonumber \\
-{4(32\ls+4\l+1) \over r^2(4\l+1)^2}(\gr-s)-{2 \over r^2(\l+1/4)}s_1- \nonumber \\
-{4 \over r^2(\l+1/4)^2}\dP-{8(32\ls-12\l-3) \over r^2(4\l+1)^2}\dB=0,
\eeq
\beq
\label{s1eq}
(h\square+r^{-5}\p_rr^5\p_r)s_1+{1 \over r(\l+1/4)}\p_r(\g+s-\gr)+{32\l \over r(4\l+1)}\p_r\dB- \nonumber \\
-{4(112\ls+8\l+3) \over r^2(4\l+1)^2}s_1-{64\l(4\l-1) \over r^2(4\l+1)^2}s_2- \nonumber \\
-{16(12\l+1) \over r^2(4\l+1)^2}\gr+{32(32\ls+4\l+1) \over r^2(4\l+1)^2}\dB=0,
\eeq
\beq
\label{s2eq}
(h\square+r^{-5}\p_rr^5\p_r)s_2+{1 \over 4r(\l+1/4)}\p_r(\g+s-\gr)- \nonumber \\
-{8\l \over r(4\l+1)}\p_r\dB-{16(4\ls-\l) \over r^2(4\l+1)^2}s_1- \nonumber \\
-{4(64\ls+28\l+5) \over r^2(4\l+1)^2}s_2-{4(4\l-1) \over r^2(4\l+1)^2}\gr+ \nonumber \\
+{8(32\ls-4\l-1) \over r^2(4\l+1)^2}\dB+{16\dP \over r^2(4\l+1)}=0.
\eeq
The 5-dimensional graviton $\g_{ij}$ includes 5 unphysical degrees of freedom associated with the (linearized) gauge transformations
\beq
\label{5gauge}
\g_{ij} \rightarrow \g_{ij}+D_i\xi_j+D_j\xi_i,
\eeq
where $\xi_i$ is an arbitrary 5-dimensional vector. In choosing $\g_{\m r}=0$ in the ansatz (104), we used 4 of these 5 degrees of freedom, so we still have to make one more gauge choice before solving the equations of motion. We find it convenient to choose the gauge 
\beq
\label{ggauge}
\p_r\g=0,
\eeq
so that the trace $\g$ of the 4-dimensional graviton is $r$-independent and completely determined by its boundary value ${\hat \g}$. With this gauge choice, the graviton $\g_{\m\n}$ decouples from the scalar equations (\ref{dileq}-\ref{s2eq}). The solutions of these equations then enter into the graviton equations (\ref{geq},\ref{phieq}) through the quantities $V,X$ defined in (\ref{Vdef},\ref{xdef}). Using eqs. (\ref{dileq}-\ref{s2eq}), we find that $V,X$ satisfy the equations
\beq
\label{Veq}
\p_r({r^4 \over \l+1/4}V)=2L^4\square\lgroup{1 \over 4}\p_r(s-\gr)+ \nonumber \\
+{2 \over r}s+{2 \over r}{\l \over \l+1/4}\dB+{1 \over 8r}{12\l+5 \over \l+1/4}(s-\gr)\rgroup,
\eeq
\beq
\label{Xeq}
\lgroup\p_r^2+({9 \over r}-{1 \over r(\l+1/4)})\p_r+({15 \over r^2}-{5 \over r^2(\l+1/4)})\rgroup X= \nonumber \\
=h\square\lgroup -{3 \over 4}\p_r\gr-{1 \over 4}\p_r s- \nonumber \\
-{1 \over 2r}{12\l+5 \over 4\l+1}\gr-{1 \over 2r}{4\l-1 \over 4\l+1}s+{8\l \over r(4\l+1)}\dB\rgroup.
\eeq
Looking ahead, we see that in the UV expansion, the right-hand sides of the above equations are treated as perturbations, so to 0th order we have
\beq
\label{0vx}
V_0 = A_0{\l+1/4 \over r^4}, \; \; \; X_0=A_1{\l-1/4 \over r^3}+A_2{1 \over r^5}.
\eeq  
Substituting the above into eq. (\ref{geq}), and using again the gauge choice (\ref{ggauge}), we find
\beq
\label{A0value}
A_0 = L^4(k^2{\hat \g}-k_{\m}k_{\n}{\hat \g_{\m\n}}),
\eeq
so $A_0$ is completely determined by the boundary data. We see that to find the leading order 2-point function, we need to solve the eqs. (\ref{dileq}-\ref{s2eq}) to first order in the UV expansion. The first order solutions will have 2 arbitrary constants which we can express in terms of $A_0$ and $A_2$. Then plugging the 0th order part of the first order solutions into the right-hand-side of eq. (\ref{Xeq}), we will solve for $X$ to 1st order, and choose $A_2$ so as to eliminate the leading order $log$ term in the IR limit. In solving eqs. (\ref{dileq}-\ref{s2eq}), the boundary conditions (\ref{ttboundary}) mean that all scalar fields should approach 0 at $r\rightarrow\infty$. The first order solutions with these boundary conditions, and with (\ref{0vx},\ref{A0value}) are:
\beq
\label{ttscalaruvsols}
\gr=-{A_0\l \over 6r^2}-{A_2 \over 4r^4}, \; \; \dB={A_0 \over 8r^2}+{A_2 \over 2r^4}, \nonumber \\
s_1={7A_0 \over 36r^2}+{A_2 \over 3r^4}, \; \; s_2={13A_0 \over 144r^2}+{5A_2 \over 12r^4}, \; \; \dP=-{A_0 \over 6r^2}+{13A_2 \over 12r^4},
\eeq
where we have only kept the leading $log$ terms. Plugging these into (\ref{xdef}), we find
\beq
\label{A1value}
A_1 = -{1 \over 2}A_0.
\eeq
Substituting the solutions (\ref{ttscalaruvsols}) into the right hand side of (\ref{Xeq}) and going to next order, we obtain
\beq
\label{1x}
X = -A_0{\l-1/4 \over 2r^3}-{5 \over 24}A_0L^4k^2{\ls \over r^5}+{A_2 \over r^5},
\eeq
where we have again kept only leading logs. From this we see that in taking the usual IR limit, we will need to choose
\beq
\label{A2value}
A_2 \sim A_0L^4k^2\log^2(k/\Lambda) = L^8k^4\log^2(k/\Lambda)({\hat \g}-{k_{\m}k_{\n}{\hat \g}_{\m\n} \over k^2}).
\eeq
As in the previous section, the above considerations only allow us to determine the coefficient $A_2$ up to a constant factor.

To complete the calculation of the EM tensor 2-point function, we now have to substitute the above solutions into the SUGRA action. We only need to be concerned with the gravitational part of the action; this is because our boundary conditions stipulate that all scalar fields (except $\g$ and $\f$) approach 0 at $r\rightarrow\infty$, so all contributions to the 2-point function from nongravitational parts of the action will vanish. The quadratic gravitational action is (see e.g. \cite{af}):
\beq
\label{quadaction}
S \sim {1 \over \kappa^2}\int d^4xdr{\sqrt g}(D_Kh_{MN}D^Kh^{MN}-2D_Mh_{KN}D^Kh^{MN}+2D_Nh_K^{\; K}-D_Mh_K^{\; K}D^Mh_N^{\; N})
\eeq
Integrating by parts in the usual way, and using eqs. (\ref{ttansatz},\ref{gammabardef},\ref{phidef}), this becomes
\beq
\label{boundaction}
S \sim {1 \over \kappa^2}\int_{r=\infty}d^4x\, r^5\lgroup{\bar \g}_{\m\n}\p_r{\bar \g}_{\m\n}+({4 \over 3}{\p_{\r}\p_{\s}{\hat \g}_{\r\s} \over \square}-{1 \over 3}{\hat \g})\p_r\f-{\hat \g}\p_r(2\gr+s)\rgroup
\eeq
Substituting eqs. (\ref{phieq},\ref{ttscalaruvsols},\ref{1x},\ref{A2value}) the terms conspire to add to a transverse momentum space 2-point function
\beq
\label{totaltt2point}
\langle T_{\m\n}(k)T_{\r\s}(-k)\rangle ={\d ^2S \over \d{\hat \g}_{\m\n}\d{\hat\g}_{\r\s}} \sim g_s^2M^4k^4\lgroup\pi_{\m\n\r\s}(k)\log^3(k/\Lambda)+C\pi_{\m\n}(k)\pi_{\r\s}(k)\log^2(k/\Lambda)\rgroup,
\eeq
where the projectors $\pi_{\m\n}$ and $\pi_{\m\n\r\s}$ have been defined in eqs. (\ref{transproj},\ref{ttprojector}) and we are unable to determine the numerical value of the constant $C$. Note that by multiplying eq. (\ref{totaltt2point}) by $\d_{\m\n}\d_{\r\s}$, we find
\beq
\label{tracettracet}
\langle T_{\m}^{\; \m}(k)T_{\m}^{\; \m}(-k)\rangle \sim g_s^2M^4k^4\log^2(k/\Lambda).
\eeq
Thus the leading order logarithmic behavior of the 2-point function of the trace anomaly scalar $T_{\m}^{\; \m}$ is different from that of the minimal massless scalar discussed in section 3, but the same as that of the R-anomaly scalar $\p\c J^R$ discussed in the previous section (see eq. (\ref{divjrdivjr})). 

\section{Discussion}

Let us restate our main results. To leading order at high energies, we find the following momentum space 2-point functions for the R-current and the energy-momentum tensor:
\beq
\label{discjj}
\langle J^R_{\m}(k)J^R_{\n}(-k)\rangle=g_s^2M^4k^2\lgroup A_0\pi_{\m\n}(k)\log^3(k/\Lambda)+B_0{k_{\m}k_{\n} \over k^2}\log^2(k/\Lambda)\rgroup,
\eeq
\beq
\label{disctt}
\langle T_{\m\n}(k)T_{\r\s}(-k)\rangle = g_s^2M^4k^4\lgroup C_0\pi_{\m\n\r\s}(k)\log^3(k/\Lambda)+D_0\pi_{\m\n}(k)\pi_{\r\s}(k)\log^2(k/\Lambda)\rgroup,
\eeq
where $A_0,B_0,C_0,D_0$ are $k$-independent constants. The most general structure of the correlators allowed by the symmetries is 
\beq
\label{genjj}
\langle J^R_{\m}(k)J^R_{\n}(-k)\rangle=A(k^2)\pi_{\m\n}(k)+B(k^2){k_{\m}k_{\n} \over k^2},
\eeq
\beq
\label{gentt}
\langle T_{\m\n}(k)T_{\r\s}(-k)\rangle =  C(k^2)\pi_{\m\n\r\s}(k)+D(k^2)\pi_{\m\n}(k)\pi_{\r\s}(k),
\eeq
where $A,B,C,D$ are $k$-dependent form factors. The presence of nonzero form factors $B$ and $D$ indicates, respectively, the anomalous breaking of R and conformal symmetries. Indeed, the  longitudinal part of the $\langle JJ\rangle$ correlator should be proportional to the R-symmetry anomaly $\p_{\m}J_{\m}$, and the trace part of the $\langle TT\rangle$ correlator should be proportional to the trace anomaly $T_{\m}^{\; \m}$. Moreover, supersymmetric Ward identities are expected to relate $A$ to $C$ and $B$ to $D$ \cite{f3}. From the functional form of the form factors in eqs.(\ref{discjj},\ref{disctt}) (the identical powers of the leading order logarithm in $A$ and $C$ and in $B$ and $D$), it seems plausible that these identities are indeed satisfied. This provides a qualitative check on our results. Note that although the $\langle JJ\rangle$ and $\langle TT\rangle$ correlators are related to each other by Ward identities, the sectors of SUGRA fluctuations that are dual to them are completely decoupled from each other; there is no interaction between the modes considered in section 4 and those in section 5. In this, the KT background is similar to AdS, and markedly different from the RG flows studied in refs. \cite{f1,f2,f3}. The reason for this is unbroken chiral symmetry. In the above papers, it is the breaking of chiral symmetry, which is an IR phenomenon, that mixes the $T$ and $J$ modes. If we were interested in the IR behavior of the same correlators in the full KS background, and not only in its UV limit -- the KT background -- we would encounter the same kind of mixing. Unlike the RG flows backgrounds, we see little hope of obtaining analytic results in the KS background.

In all the 2-point functions we have computed, we encounter at leading order logarithmic factors of the form $\log^n(k)$, where $n$ is a positive integer (an exponent of $n$ in momentum space translates to an exponent of $n-1$ in position space). From the way these logarithms arise in the UV solutions to the fluctuation equations, it is easy to see that in general, the larger the dimension of the field theory operator $\cal{O}$ whose 2-point function $\langle\cal {OO}\rangle$ we are calculating, the more times we need to iterate the UV expansion, and since in each iteration we effectively pick up a factor of $g_sM\log(k)$, the higher the power of the logarithm that will appear in the 2-point function. The interpretation of these logarithms from the field theory point of view is somewhat mysterious. On the one hand, the $SU(N+M)\times SU(M)$ gauge theory has a nontrivial beta function, with the relative coupling flowing as
\beq
\label{beta}
{1 \over g_1^2}-{1\over g_2^2} \sim M\log(\m),
\eeq
where $\m$ is the energy scale; the Yang-Mills coupling $1/g_s \sim 1/g_1^2+1/g_2^2$ remains constant in the supergravity approximation \cite{ks}. Thus a perturbative expansion in the coupling would be an expansion in inverse powers of the logarithm, but it is difficult to see how it could give rise to the large positive powers expected to appear for operators of large dimension. It is tempting to speculate that the logarithmic growth of the correlators is a manifestation of a logarithmic growth with scale of the effective number of degrees of freedom in the theory. Indeed, as noted above, the logarithms always appear in the combination $M\log(k)$ which perhaps somehow represents the ``effective number of colors'' in the UV. Some support for this is provided by the fact that when the finite temperature theory is considered, one finds an entropy that grows logarithmically with scale \cite{buchel}. It would be interesting to better understand the field theory origin of these logarithmic factors.

\acknowledgments

I am grateful to Igor Klebanov for suggesting these problems to me, and for many important discussions. I am also indebted to Kostas Skenderis for alerting me to the relevance of holographic renormalization in the present context. I thank Peter Ouyang for useful remarks.

\end{document}